\documentstyle[preprint,aps,prd]{revtex}
\begin{document}
\draft
\title{Mixed dark matter with low-mass bosons}
\author{Georg B.~Larsen and Jes Madsen}
\address{Theoretical Astrophysics Center and
Institute of Physics and Astronomy, University of Aarhus, 
DK-8000 \AA rhus C, Denmark}
\date{September 29, 1995}
\maketitle

\begin{abstract}
We calculate the linear power spectrum for a range of mixed dark matter
(MDM) models assuming a massive (few eV) boson, $\phi$,
instead of a neutrino as the
hot component. We consider both the case where the hot dark matter (HDM)
particle is a boson and the cold component
is some other unknown particle, and the case where there is only one
dark matter particle, a boson, with the cold dark matter (CDM) component 
in a Bose condensate. Models resembling the latter type could arise from
neutrino decays---we discuss some variants of this idea.
The power spectra for MDM models with massive bosons 
are almost identical to neutrino MDM models for a given mass fraction of HDM
if the bosons are distinct from their antiparticles ($\phi\neq\bar\phi$)
and have a temperature like that of neutrinos, whereas models with
$\phi=\bar\phi$ tend to overproduce small-scale structure.
\end{abstract}

\pacs{95.35.+d, 14.80.-j, 98.65.-r, 98.80.-k
$\,\,\,\,$ (To appear in Phys. Rev. D15 March)}

\section{Introduction}
The standard CDM model overproduces structure in the
Universe on small and intermediate length scales (up to 10--30 Mpc)
when normalized to the large-scale fluctuations in the 
cosmic microwave background radiation measured by COBE \cite{cobe}. The mixed 
dark matter model\cite{shafi} has been one of the most successful 
modifications of standard CDM.
When normalized to the COBE data it can reproduce the
right amount of structure in the linear regime\cite{peacock}, 
and fit the observations on galaxy scales\cite{shafi}.
For many people MDM models are unattractive in that they
involve two different types of dark matter particles, the neutrino with an eV
mass and some other CDM particle. It was shown by Madsen \cite{madsen1}
that if a heavy fermion (perhaps a neutrino) decays
into a fermion and a boson in the early Universe, then a large fraction of the
bosons can be formed in a Bose condensate (CDM) while the rest of the bosons
are thermal (HDM). Kaiser, Malaney \& Starkman\cite{kaiser} subsequently 
demonstrated that one indeed gets a hot as well as a very
cold component, though not actually a condensate. They dubbed the
formation process ``neutrino lasing''.
This could be a physical explanation for MDM models and only
one dark matter particle is needed, a boson with a mass of a few eV.

Motivated by the ideas outlined above, this paper studies MDM models with eV
mass bosons rather than neutrinos as hot dark matter. 
First we discuss the fundamental differences between bosons and fermions
in the context of structure formation (for pure HDM this was originally
studied in\cite{madsen2}). Then we describe the method used to
compute the linear power spectrum, and show our results compared to the 
observed linear power spectrum. Finally we draw our conclusions
based on how well a given model fits the observed amount of structure today,
also including some comments on damped Lyman-$\alpha$ systems 
(DLAS) to constrain the models on small (galactic) scales.
We study variants of the model where the bosons are in kinetic
equilibrium, and also comment on the consequences of relaxing this
assumption. The models give structure
formation results that are virtually identical to neutrino MDM for fixed
HDM fraction if the
boson, $\phi$, is not its own antiparticle, whereas models with
$\phi=\bar\phi$ tend to overproduce small-scale
structure (assuming the temperature of $\phi$ equal to the neutrino
temperature). For the preferred mixing ratio calculated from the
decay scenario, the small-scale power is inconsistent with observations.

\section{Bosons as dark matter}

The momentum distribution function for ultrarelativistic particles in 
kinetic equilibrium
is given by:
\begin{equation}
f(p)={1 \over {\exp ({p-\mu\over T}) \pm 1}},
\label{ferbos}
\end{equation}
where $T$ is the temperature and $\mu$ is the chemical potential of the 
species. The $+1$ corresponds to a Fermi-Dirac distribution and $-1$ to a 
Bose-Einstein distribution. 

The consequences of a non-zero chemical potential 
for neutrino MDM models were discussed
in \cite{larsen}. If the density 
of relativistic particles is increased (roughly by a
factor of 2) then MDM and also CDM models give a much better fit
to the observed linear power spectrum. A non-zero chemical potential is a simple
way to accomplish that.
Here we assume the chemical potential to be zero. In the case of bosons this 
allows for the existence of a Bose condensate if the bosons are formed with a
temperature below a critical temperature $T_c=(\pi^2n_B/\zeta(3))^{1/3}$,
where $n_B$ is the number density of bosons\cite{madsen1}.

The hot bosons are assumed to be thermally distributed with a number density
calculated from Eq.\ (\ref{ferbos}) with a temperature $T_\phi$. 
As shown in \cite{kaiser} the hot
component in decay models is not thermal, but peaked in momentum space.
However, for MDM models the important quantities (for a fixed HDM mass
fraction, $\Omega_{\rm HOT}$) are the mean momentum of the hot
component and the particle mass rather than the detailed distribution,
so use of Eq.\ (\ref{ferbos}) is a fairly good approximation.
We shall return to this question later.

Given a certain mass fraction of HDM the mass of the HDM particle is
fixed. In standard MDM models
one of the three neutrino species is assumed to have a mass given by
\begin{equation}
m_\nu=93.8{\rm eV}\Omega_\nu h_0^2,
\end{equation}
i.e.\ $m_{\nu}=4.7$eV
for a Hubble parameter of $h_0=0.5$, a neutrino mass fraction
$\Omega_\nu =0.2$, and a neutrino temperature $T_\nu=(4/11)^{1/3}T_0$,
where the COBE measured 
temperature of the cosmic microwave background radiation is $T_0=2.726$K 
\cite{mather}. For $T_\phi=T_\nu$ the mass of the boson
will be a factor of 1.5 (or 0.75) times the mass of a neutrino for the same
hot mass fraction $\Omega_{\phi,{\rm HOT}}=\Omega_{\nu}$ 
depending on whether or not $\phi=\bar\phi$, since in general
\begin{equation}
m_\phi=140.7{\rm eV}\Omega_{\phi,{\rm HOT}}h_0^2 g^{-1}\left(\frac{T_\nu}
{T_\phi}\right)^3 ,
\label{thermal}
\end{equation} 
where $g=1$ (2) for $\phi=\bar\phi$ ($\neq\bar\phi$).
In most of the paper we assume $T_\phi=T_\nu$, corresponding to a
particle in thermal equilibrium decoupling between the QCD phase transition
at $T\approx 100$MeV and electron-positron annihilation at $T\approx
0.5$MeV, but we shall comment on other possibilities later.
Notice that the
boson mass is determined by the thermal (hot) component alone. If bosons
exist also in a cold (condensate) component there is an extra
contribution to the number density of bosons given by $\Omega_{\phi,{\rm
COLD}}/\Omega_{\phi,{\rm TOTAL}}$. 

In order to be consistent with the constraints from Big Bang nucleosynthesis
(BBN) we assume only 2 massless neutrino species and one boson species (low
mass, thus ultra-relativistic at BBN). The third neutrino species is assumed to
have decayed away before BBN. The effective number of neutrino families at the
epoch of BBN is then (for $T_\phi=T_\nu$) 
$N_{\rm eff}=2.57$ (3.14) if $\phi=\bar\phi$ 
$(\phi\neq\bar\phi)$, giving nucleosynthesis predictions within the
observationally allowed range\cite{smith}. We shall comment later on
the effects of having an extra massless neutrino flavor.

\subsection{Qualitative effects on structure formation}

The amount of power in the density perturbation spectrum erased by 
free streaming of hot particles depends on the {\it rms} 
velocity of the massive bosons through the Jeans wavenumber given by:
\begin{equation}
k^2_{\rm J,HOT}\equiv {4\pi G\rho_0a^2 \over v^2_{rms}},
\end{equation}
where $a$ is the scale factor and $\rho_0$ is the critical density (we assume
$\Omega_0=1$ throughout this paper).
All the power of the hot dark matter with a wavenumber greater
than the Jeans wavenumber at $t_{\rm eq}$ 
(the time when the Universe shifts from being dominated by radiation 
to matter domination) will have free streamed away.
The Jeans wavenumber 
grows proportional to $a^{1/2}$ 
(the velocity decreases as $v_{rms}\propto a^{-1}$ and the density goes as
$\rho_0\propto a^{-3}$ as the 
Universe expands). Only when the Jeans wavenumber has
become larger than the wavenumber of a given HDM density perturbation, 
that perturbation can begin to grow again.
The growth rate of the cold component is suppressed because of the 
more homogeneous hot
component. The shape of the power spectrum is determined by the Jeans 
wavenumber at $t_{\rm eq}$ and the fraction of HDM,
$\Omega_{\rm HOT}$.  The Jeans wavenumber determines where the 
MDM spectrum breaks away from CDM, and 
$\Omega_{\rm HOT}$ determines the bending of the power spectrum. 

The {\it rms} velocity decisive for $k_{\rm J,HOT}$ 
can be calculated from Eq.\ (\ref{ferbos}).
Free streaming is most severe when
particles are relativistic and $v_{rms}=c$. For typical HDM
particle masses particles become nonrelativistic just before or 
around $t_{\rm eq}$, after which the velocity decreases and free
streaming becomes less important.

The main differences between having a boson HDM particle or a neutrino 
(fermion) are in the masses of the HDM particle, and in the
different velocity distributions. The nonrelativistic 
{\it rms} velocity depends on the mass 
of the particle, $v_{rms}\propto m^{-1}$, and also on the phase-space 
distribution. The two differences happen to cancel out each other in the 
boson MDM models with $\phi\neq\bar\phi$. The mass $m_{\phi}$ is a factor 0.75 
lower than in the neutrino MDM model, but the more low momentum states of
bosons relative to neutrinos compensates this, and makes the two power 
spectra look very much the same, c.f.\  Fig.\ 1.

\section{Numerical results and discussion}

In order to calculate the linear power spectrum it is necessary to integrate
the linearized equations of general relativity and the Boltzmann equation for
the HDM particles. We use the program package of Bertschinger, 
COSMICS \cite{bertschinger}, which
can integrate the linearized equations in both the synchronous and conformal
Newtonian gauge. We have modified the FORTRAN program linger-syn.f
(synchronous gauge) changing the phase space distribution function and making 
some other appropriate changes, so that we could calculate the linear power
spectrum using bosons as hot dark matter. As initial conditions we 
have assumed adiabatic density perturbations with an initial power spectrum
of Harrison-Zel'dovich type, $P(k)\propto k^n$ and $n=1$, as predicted by
most theories of inflation. 

In all the MDM models calculated we have assumed $\Omega_0=1$ and in
most models a Hubble parameter of $h_0=0.5$.
Each power spectrum was calculated with $40$ wavenumbers ranging from 
$k=10^{-5}$ Mpc$^{-1}$ to $k=2$ Mpc$^{-1}$, and then expanded to 201 points
using the program grafic.f \cite{bertschinger}. The power spectrum was
normalized using the COBE measured value of the microwave background anisotropy
of $Q_{rms-PS}=17\pm 3\mu$K \cite{cobe}. The temperature of the microwave 
background radiation is taken to be $T_0=2.726$K today \cite{mather}. We have
throughout assumed 5\% baryons, $\Omega_{\rm baryon}=0.05$.

When the calculated linear power spectrum is to be compared with the 
observations of large scale structure in the Universe we have chosen to 
compare with the reconstructed linear power spectrum given by Peacock and Dodds
\cite{peacock}. They have combined several surveys of different types of 
galaxies and clusters, and then corrected for redshift distortions and
non-linear effects. While this reconstruction is somewhat model
dependent\cite{white}, and it would be better to compare the actual data to
simulations in the non-linear regime, it is expected to give a
reasonable discrimination between models. It is important to bear in mind,
though, that the error bars in Ref.\ \cite{peacock} almost certainly 
are too small because the averaging procedure used underestimates the
systematic errors that may result from the correction for 
bias between the different samples. If, for example, the IRAS galaxies are 
biased by a factor $b_I$ less than one relative to the dark matter that would 
raise the power spectrum and give a better fit to the calculated curves 
in the Figures below. From Ref.\ \cite{peacock} the bias can be as low as
$b_I=0.8$ which would raise the data points by a factor $b_I^{-2}=1.5$. 
Corrections for non-linear evolution are
particularly important for large $k$, and could lead to systematic
errors for the last few data points shown in our Figures, but these
errors are probably minor (less than a factor of 2), 
as the region shown is only mildly non-linear. 

In spite of all the reservations, we will use
the data points from Peacock and Dodds\cite{peacock} to guide the eye. The main
conclusions about the quality of our models can be made by comparing 
the boson MDM power spectra with the
standard MDM power spectrum which is known to make a good fit for 
more detailed comparisons with observations\cite{shafi}.

In order to compare directly with the data points of Peacock and Dodds we 
follow their notation and calculate the dimensionless power spectrum,
\begin{equation}
\Delta^2(k)\equiv {k^3P(k) \over 2\pi^2},
\end{equation}
which we henceforth will refer to as the power spectrum. $\Delta^2(k)$ can be 
described as the contribution to the fractional density variance per 
logarithmic interval in $k$.

\subsection{Thermal models}

\subsubsection{$T_\phi=T_\nu$}

In Fig.\ 1 we have shown the power spectra of three different MDM models. The
mass fraction of hot dark matter varies with four values $\Omega_{\rm HOT}=0.15,
 0.20, 0.25, 0.30$, from the top curve to the bottom. The standard MDM models
with one massive neutrino species and two massless are plotted as the 
dash-dotted curves. This should be directly compared to the MDM models with
one massive boson $(\phi\neq\bar\phi)$ and two massless neutrino species, the
solid curves. Only a very small difference between the power spectra of the 
two models is evident. This is due to the cancellation effect mentioned earlier.
The less massive bosons (a factor of 0.75 relative to the neutrino mass for
$\phi\neq\bar\phi$) lead to a higher value of the {\it rms} velocity
giving more free streaming. But this is compensated by the difference in the 
phase-space distribution function, with more bosons found in the low momentum 
states relative to neutrinos giving less free streaming. 

The dashed curves in Fig.\ 1 are the corresponding boson MDM models with 
$\phi =\bar\phi$. Now (due to the higher boson mass) the Jeans wavenumber at 
$t_{\rm eq}$ is almost a factor of two larger than for neutrino MDM or
boson MDM with $\phi\neq\bar\phi$. This means that the MDM power spectrum
breaks away from the standard CDM, dotted curve, at a wavenumber a factor of
two larger, resulting in a somewhat worse fit to the data points of large scale
structure. Notice that the dashed curves bend in the same way as the other
power spectra, because the mass fraction of HDM takes the same four values
for all MDM models in this figure (the growth rate of the CDM component is
reduced by the same amount).

When comparing with the observed linear power spectrum, it is clear that the
standard CDM model when normalized to the COBE data makes a very poor fit.
The MDM models with $\phi =\bar\phi$ break away from CDM a bit too late, 
that is at too large wavenumbers, giving a somewhat poor fit, but 
not excluded in case of a lower normalization on COBE-scales, for
example due to gravity wave contributions to the large scale anisotropy.
The standard MDM models are seen to fit the data
points pretty well, apart from a constant factor which can be accommodated by
lowering the normalization or by introducing a bias parameter. The COBE 
normalization of the power spectrum depends on the quadrupole
$Q_{rms}$ squared, so within
the errors the curves can be lowered by a factor of $1.4$. And as discussed
above a bias of $b_I=0.8$ would raise the data points by a factor $1.5$.
Rather than tuning these parameters to get better fits in our Figures,
we define a model as being in reasonable agreement with observations if
it has a power spectrum similar to that of the well-tested standard MDM
model \cite{shafi}.

The favored value of the hot fraction is 
$\Omega_{\rm HOT}\simeq 0.20$--0.25. 
The same is true for MDM models with a massive
boson $(\phi\neq\bar\phi)$ and two massless neutrino species.

\subsubsection{$T_\phi\neq T_\nu$}

The discussion above assumed thermal bosons with a temperature equal to
that of neutrinos in the standard model. Clearly, relaxing this
assumption changes the power spectra. For a fixed $\Omega_{\phi,{\rm
HOT}}$ a decrease in $T_\phi$ (for example due to boson decoupling prior
to the QCD phase transition) requires a more massive boson, so that the
power spectra will follow standard CDM over a wider range of wave
numbers before breaking away. This clearly makes the models less
attractive compared to observations. On the other hand a hotter boson
component could be more attractive, since it would be less massive and
free stream more efficiently. Such a situation would be hard to obtain
for a primordial, thermal boson (it could come about only if decoupling
took place after electron-positron annihilation, which would increase
its abundance during BBN, and also require probably unrealistically strong
interactions). Something resembling this could, however, be an outcome
of the decay scenario\cite{madsen1,kaiser}.

\subsection{Decay models}

The physically interesting models which allow MDM with only one 
dark matter particle, a massive boson, tend to predict a high fraction of
hot dark matter in the Universe\cite{madsen1,kaiser}. 
When the cold component of 
bosons formed during stimulated decay of e.g.\ a heavy neutrino, only half
of the heavy neutrinos decay in this way, the rest of them decay into the hot
bosons (and fermions). The resulting fraction of cold bosons is only 35\% in 
the case of $\phi\neq\bar\phi$ \cite{kaiser}. We have shown the power
spectrum of such an MDM model 
with $\Omega_{\rm HOT}=0.6$, $\Omega_{\rm COLD}=0.35$ and 
$\Omega_{\rm baryon}=0.05$ as the lower solid curve in Fig.\ 2.

The main problem with MDM models having 60\% hot dark matter is that the 
power spectrum bends too much at small scales (high values of $k$).
The linear power spectrum
of this MDM model has most power on scales of $k\simeq 0.08$Mpc$^{-1}$ and less
power on smaller scales, which would make it inconsistent with hierarchical 
structure formation. In Fig.\ 2 is also shown the boson MDM model with a
hot fraction of $\Omega_{\rm HOT}=0.25$ the top solid curve. This model 
fits the large scale structure data points as well as any standard MDM
model, but it assumes
twice as many bosons in a Bose condensate as predicted by neutrino lasing.

So far we have assumed a boson mass given by the thermal distribution,
Eq.\ (\ref{thermal}). As mentioned earlier, neutrino lasing in fact does
not lead to a thermal component, so in this case we should instead
calculate the mass from the actual number density, which corresponds to
that of the fermion decaying into the boson plus
the small number density of primordial thermal bosons.
The mass of the boson would then be significantly higher
because it would roughly correspond to substitute $\Omega_{\nu}$ in Eq.\
(2) with $\Omega_{\phi}=0.95$. With this high value of the boson mass,
the Jeans wavenumber would increase considerably and the power spectrum
would break away from standard CDM too late. 
This could, however, be compensated by having a higher temperature (or,
for a non-thermal distribution, mean momentum).
As discussed in \cite{kaiser} several variations of the decay models could
perhaps allow such a possibility.

\subsection{Constraints from damped Lyman-$\alpha$ systems}

Most of the discussion above was based on the linear or weakly
nonlinear evolution of the power spectrum, i.e.\ on the large scale structure.
In order to constrain the MDM models on galactic scales ($\lambda\simeq 1$Mpc) 
where nonlinear effects are most important, one recently
discussed method is comparing with the observed number density of damped
Lyman-$\alpha$ systems (DLAS). MDM models form structure late, that is at small
redshifts compared to standard CDM, which makes it difficult to form enough 
structure at high redshifts to account for the DLAS\cite{subramanian,klypin}. 
The DLAS give constraints on the mass fraction of
collapsed baryons $\Omega_{\rm gas}$ at high redshifts. From the observations 
which are quite uncertain at high redshifts, 
the mass fraction of collapsed baryons
at $z=3-3.5$ is determined to be $\Omega_{\rm gas}=6.0\pm 2.0\times 10^{-3}$
\cite{lanzetta}. This value may be too high; using the observed DLAS from
the APM QSO survey Storrie-Lombardi {\it et al.} find a value of roughly
$\Omega_{\rm gas}=3.5\pm 1.0\times 10^{-3}$\cite{storrie} 

We calculate $\Omega_{\rm gas}$ using the Press-Schecter approximation
\cite{press} following the method outlined in \cite{klypin}. 

\begin{equation}
\Omega_{\rm gas}=\Omega_{\rm baryon}{\rm erfc}
({\delta_c\over\sqrt{2}\sigma}),
\end{equation}
where erfc is the complementary error function, $\delta_c=1.4$ from
normalizing the Press-Schecter approximation to numerical simulations
\cite{klypin}, and $\sigma=\sigma(r_f,z)$ is computed for a mass of $10^{11}$
solar masses corresponding to $r_f=0.452$Mpc using Gaussian smoothing 

\begin{equation}
\sigma^2(r_f,z)=\int\Delta^2(k)W^2(kr_f){dk\over k},
\end{equation}	
where the window function is a Gaussian, $W^2(kr_f)=\exp(-(kr_f)^2)$ and the
power spectrum has to be calculated out to greater wavenumbers $k=10$Mpc$^{-1}$
and at a redshift of $z=3.25$.

For the MDM model with $\phi\neq\bar\phi$ having only 20\% HDM
and 75\% CDM we calculate a value of $\Omega_{\rm gas}=3.3\times 10^{-3}$
(the corresponding neutrino MDM model leads to a virtually identical
result, $3.5\times 10^{-3}$, whereas boson MDM with $\phi=\bar\phi$
gives $5.6\times 10^{-3}$), which should
be enough to account for the observed amount. With 25\% HDM and 70\% CDM
the number is $\Omega_{\rm gas}=1.3\times 10^{-3}$ ($1.4\times
10^{-3}$ for neutrino-MDM, and $3.2\times 10^{-3}$ for boson MDM with
$\phi=\bar\phi$), while having 30\% hot and 65\%
cold dark matter bosons gives $\Omega_{\rm gas}=0.36\times 10^{-3}$
($0.42\times 10^{-3}$, $1.6\times 10^{-3}$). 
These numbers indicate that only MDM models with 
$\Omega_{\rm HOT}<0.25$ can be used, a conclusion that is independent of
whether the HDM is neutrinos or bosons with $\phi\neq\bar\phi$ (for
$T_\phi=T_\nu$). Boson MDM with $\phi=\bar\phi$ is consistent with
DLAS-constraints for slightly higher $\Omega_{\rm HOT}$, but at the cost
of overproducing structure at somewhat larger length scales.
Finally we find for the MDM model with the predicted
ratio of HDM and CDM from neutrino lasing, 60\% hot 35\% cold, a very low
value of $\Omega_{\rm gas}\approx 10^{-9}$. Though the Press-Schecter 
formalism relies on hierarchical structure formation, which is not
the case when $\Omega_{\rm HOT}>\Omega_{\rm COLD}$, it demonstrates the
impossibility of forming enough small-scale structure for these parameters.

\subsection{Changing the Hubble parameter or having 3 massless neutrinos}

To show the effect on the power spectrum of having a different
value of the Hubble
parameter, we have for the boson MDM model with
30\% hot and 65\% cold dark matter and $\phi=\bar\phi$ calculated the
power spectrum for five different values of $h_0$ ranging from $0.4$ to $0.8$
in steps of $0.1$. In all models $\Omega_0=1$, so the age of the Universe
is given as $t_0=6.5h_0^{-1}$Gyr. Recent observational results have
indicated a high value for $h_0$, $h_0\simeq 0.8$\cite{pierce} which
would lead to a very low age of the Universe. From Fig.\ 3 we see that 
structure formation models work best for low values of $h_0$. The effect of
having $h_0=0.4$ instead of $0.5$ is quite big. 

In the MDM models with one massive boson species we have assumed that
one of the neutrino species had decayed away before BBN leaving only two 
massless neutrino species. When we assume $\phi=\bar\phi$ the constraints
from BBN\cite{smith} might allow for three massless neutrino species 
and the boson, the
effective numbers of neutrino families would be $N_{\rm eff}=3.57$. 
In Fig.\ 4 we have
plotted the boson MDM model with 30\% hot 65\% cold dark matter and 
$\phi=\bar\phi$ for both two and three massless neutrino species. The effect
on the power spectrum is seen to be small. The power is smaller with
more neutrinos because increasing the amount of 
radiation delays the epoch of matter
domination and suppresses the growth of structure.

\section{Conclusions}

We have demonstrated that MDM models with a low mass thermal boson
instead of a neutrino would make an equally good fit to the observed
linear power spectrum when $\phi\neq\bar\phi$. The MDM models with 
$\phi=\bar\phi$ tend to overproduce structure on small scales but not
by much, and can be made to fit the data points in \cite{peacock} by
choosing a lower value of normalization. The errors on the observed quadrupole
moment allows us to lower the curves by a factor $1.4$. It is also possible 
that some of the observed COBE
quadrupole moment might be due to gravitational radiation instead of
just the pure Sachs-Wolfe effect. With both a bias factor and a low
normalization it is possible to make the MDM models with $\phi=\bar\phi$ fit
the observed linear power spectrum, without having to introduce gravity waves.
Note that an analysis of the 2-year COBE data tend to give a higher value
of the quadrupole moment, $Q_{rms-PS}=20\pm 3\mu$K \cite{gorski}, in which 
case we would need some gravity waves. Soon the analysis of the 4-year data
will be announced and the issue of normalization hopefully settled.  

The boson MDM model predicted by lasing with 60\% hot and 35\% cold dark
matter in the form of bosons is shown to be ruled out by its inability
to account for the observed amount of small-scale structure, but other
variants of the decay scenario could be consistent with observations.
A more detailed calculation (using the actual distribution functions) would be
necessary in order to present the ``real'' power spectrum of neutrino
lasing MDM models. We believe that assuming a thermal boson
plus a Bose condensate is a good first order approximation.

\acknowledgments

We would like to thank the Theoretical Astrophysics Center under the Danish
National Research Foundation for its financial support. We thank Ed
Bertschinger for providing the COSMICS-package, which was developed under
NSF grant AST-9318185.

\begin{figure}
\caption{The dimensionless power spectrum $\Delta^2(k)$ plotted as a
function of wavenumber in units of Mpc$^{-1}$ for $h_0=0.5$ and
$\Omega_0=1$. Three different MDM models are studied, and for each model
the power spectrum for four different choices of $\Omega_{\rm {HOT}}$ are shown.
The value of $\Omega_{\rm {HOT}}$ is $0.15, 0.20, 0.25, 0.30$ 
ranging from the top
curve to the bottom. The dash-dotted curves are the standard MDM models with
one massive neutrino and two massless. The solid curves are the MDM models with
one massive boson $(\phi\neq\bar\phi)$ and two massless neutrino species.
The dashed curves represent
the corresponding MDM models with $\phi=\bar\phi$.
All power spectra are normalized to the COBE quadrupole anisotropy.
The data points are taken from \protect\cite{peacock}.
Also shown for comparison is the standard CDM power spectrum, the dotted 
curve.}
\label{fig1}
\end{figure}

\begin{figure}
\caption{The dimensionless power spectrum $\Delta^2(k)$ plotted as a
function of wavenumber in units of Mpc$^{-1}$ for $h_0=0.5$ and
$\Omega_0=1$. The power spectrum for two different MDM models 
with $\phi\neq\bar\phi$ are plotted as solid curves. The upper solid 
curve corresponds to $\Omega_{\rm HOT}=0.25$, whereas the lower
solid curve has $\Omega_{\rm HOT}=0.60$ and thus only $\Omega_{\rm COLD}=0.35$ 
(the hot/cold ratio predicted in \protect\cite{kaiser}). 
All power spectra are normalized to the COBE quadrupole anisotropy.
The data points are from\protect\cite{peacock}.
Again shown for comparison is the standard CDM power spectrum, the dotted
curve, and a neutrino MDM model with $\Omega_{\rm HOT}=0.25$
(dash-dotted curve).}
\label{fig2}
\end{figure}

\begin{figure}
\caption{The dimensionless power spectrum $\Delta^2(k)$ 
plotted as a function of wavenumber now in units of $h_0$Mpc$^{-1}$. We
have shown the power spectrum of the boson MDM model with $\Omega_{\rm HOT}=
0.3$ and $\Omega_{\rm COLD}=0.65$ (5\% baryons) and $\phi=\bar\phi$ 
calculated for five different
values of the Hubble parameter ranging from $0.4$ to $0.8$ in steps of $0.1$
from the bottom
to the top solid curve. Also plotted is the standard CDM power spectrum with
$h_0=0.5$, 
the dotted
curve.
All power spectra are normalized to the COBE quadrupole anisotropy, and
the data points are from\protect\cite{peacock}.
}
\label{fig3}
\end{figure}

\begin{figure}
\caption{The dimensionless power spectrum $\Delta^2(k)$ as a
function of wavenumber in units of Mpc$^{-1}$ for $h_0=0.5$ and
$\Omega_0=1$. All three curves assume $\Omega_{\rm HOT}=0.30$ and
$\Omega_{\rm COLD}=0.65$. 
The solid curve is the power spectrum of the boson MDM model
with $\phi=\bar\phi$ and
two massless neutrino species. The dashed curve represent the same MDM model
but now with three massless neutrino species instead of just two. To compare
the effect of the extra massless neutrino species we plot the boson MDM model
with $\phi\neq\bar\phi$ and two neutrinos as the dash-dotted curve. 
All power spectra are normalized to the COBE quadrupole anisotropy, and
the data points are from\protect\cite{peacock}.
}
\label{fig4}
\end{figure}
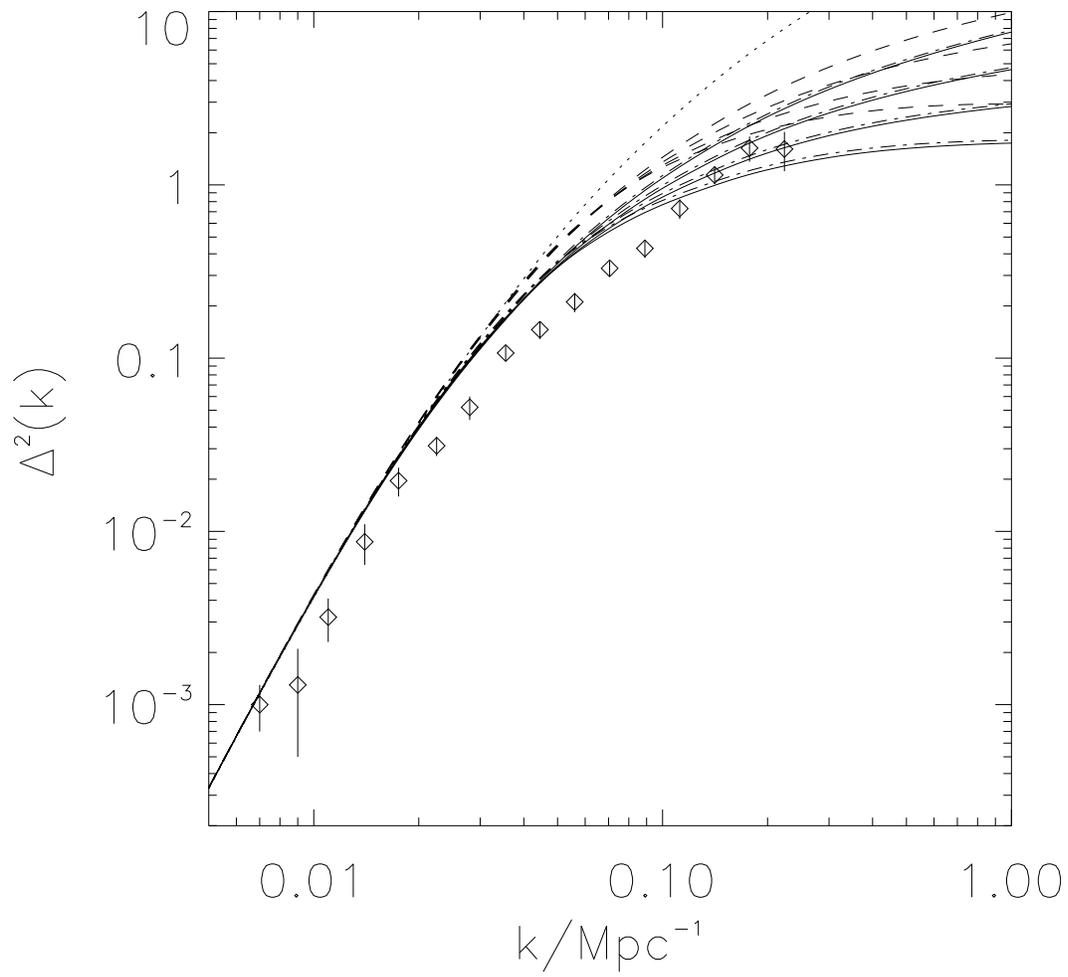
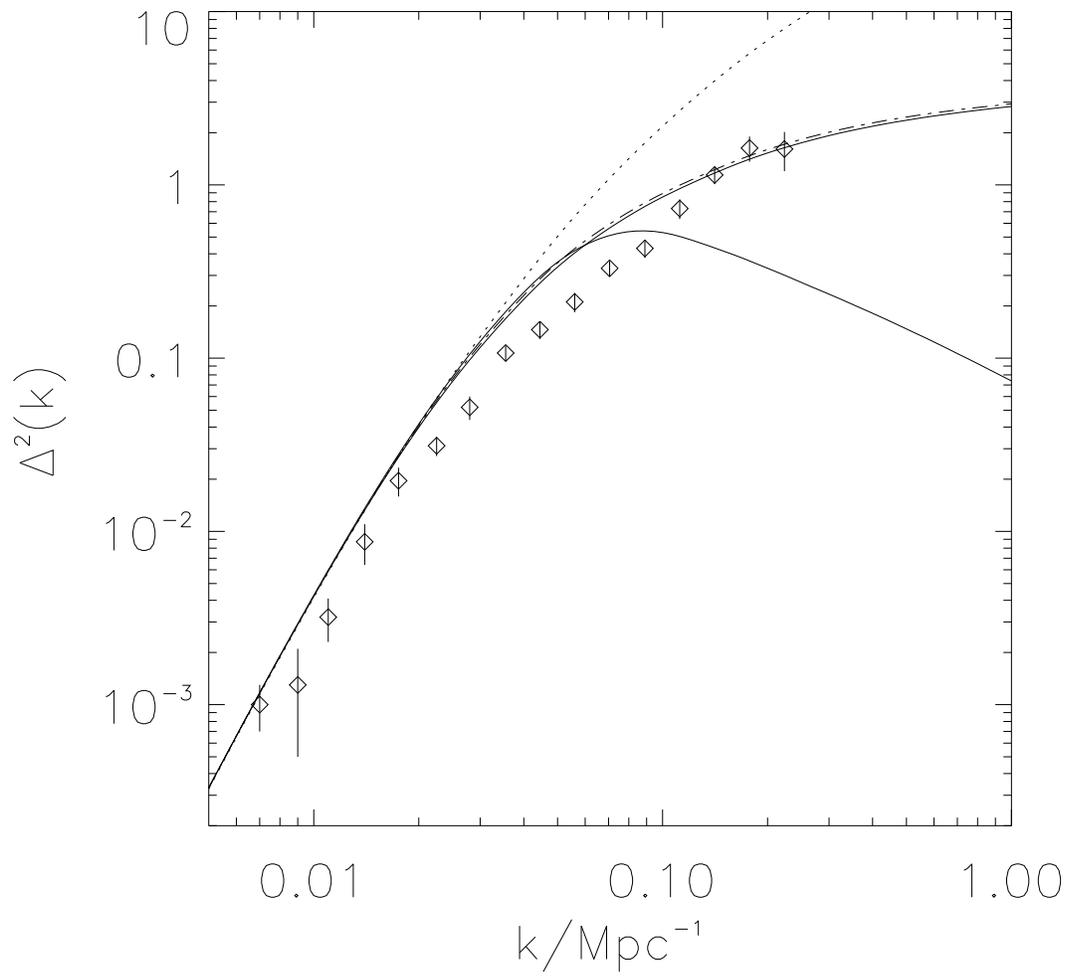
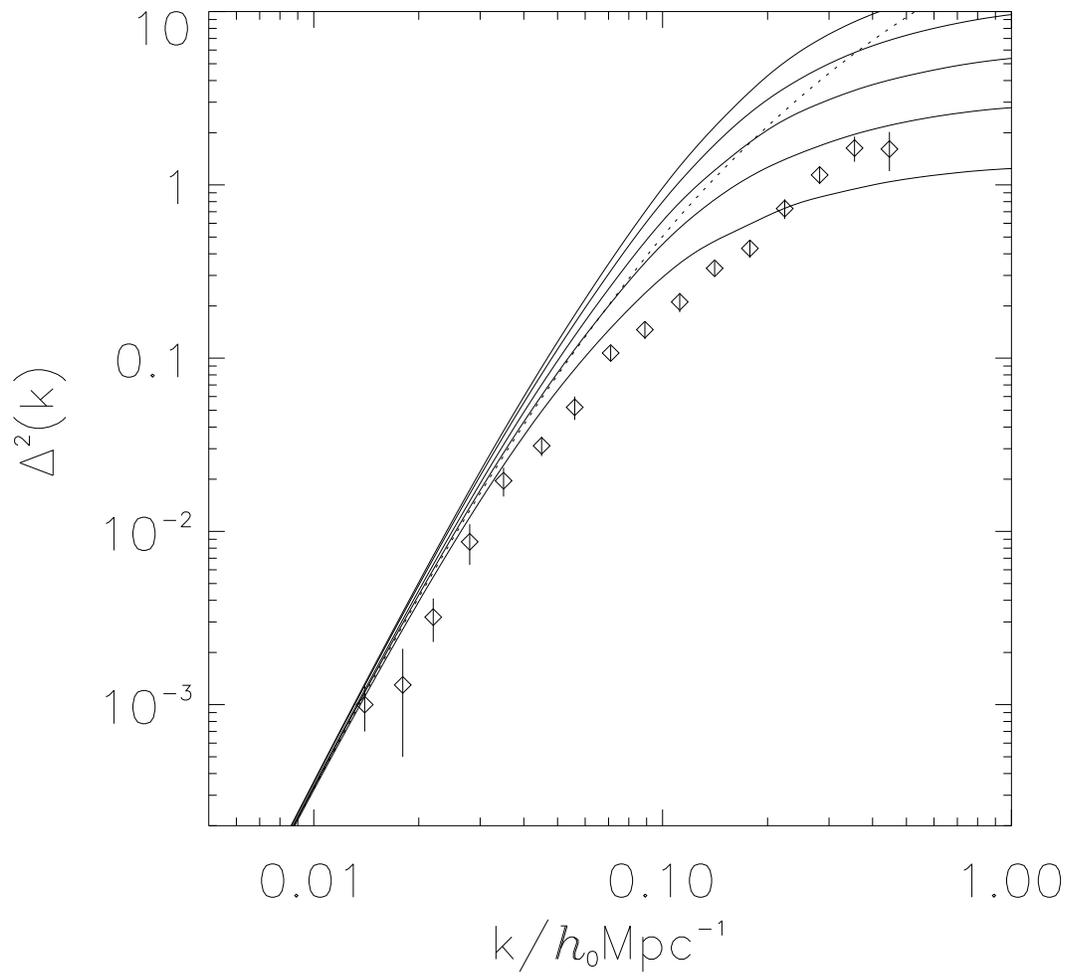
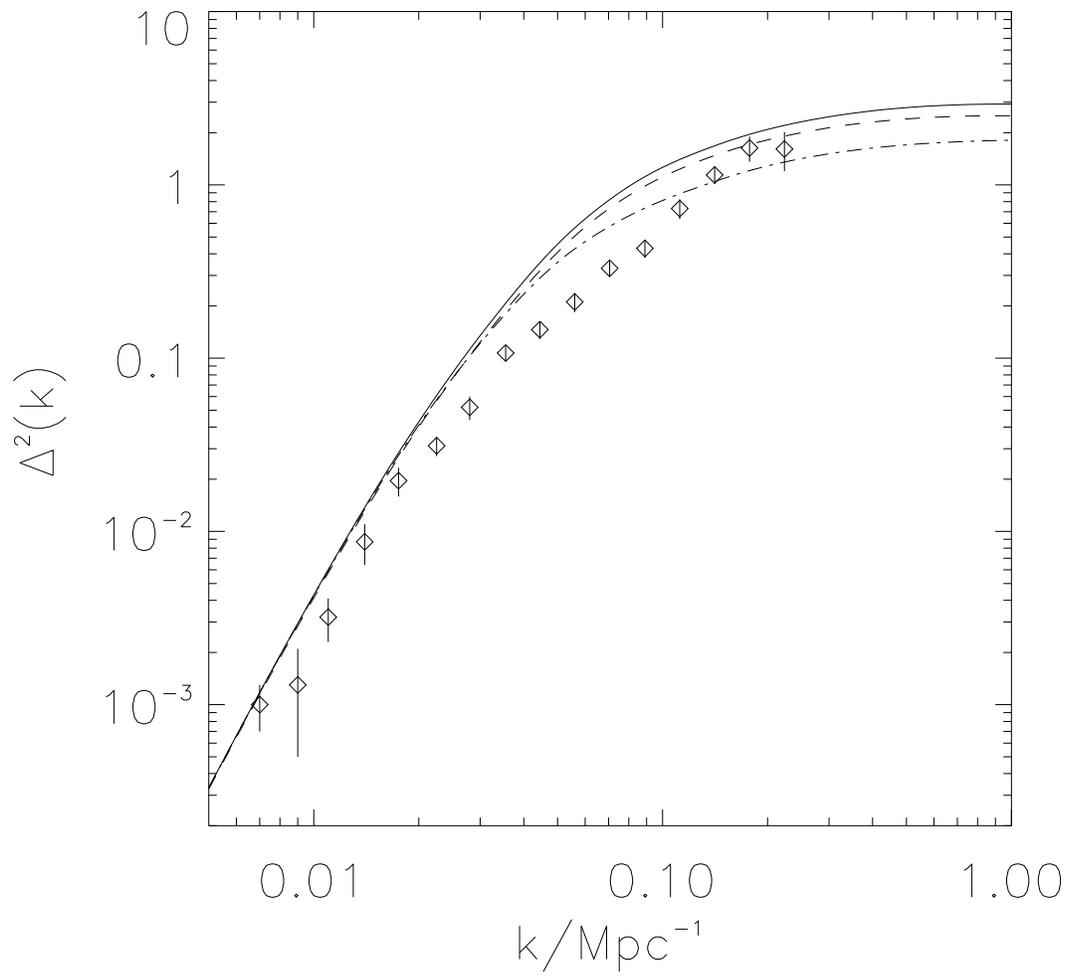
\end{document}